\documentstyle[12pt]{article}
\begin{document}
\vspace*{-1in}
\begin{flushright}
CERN-TH.7371/94 \\
TIFR/TH/94-27 \\
July 1994
\end{flushright}
\vskip 65pt
\begin{center}
{\Large \bf \boldmath $J/\psi + \gamma$ production at the Tevatron
Energy}\\
\vspace{8mm}
{D.P.~Roy$^{*}$}\\
\vspace{5pt}
{\it Theory Group, Tata Institute of Fundamental Research,\\
Homi Bhabha Road, Bombay 400 005, India.}\\
\vspace{15pt}
{and}\\
\vspace{12pt}
{K. Sridhar$^{**}$}\\
\vspace{5pt}
{\it Theory Division, CERN, \\ CH-1211, Geneva 23, Switzerland.}\\

\pretolerance=10000

\vspace{70pt}
{\bf ABSTRACT}
\end{center}
We study the process $\bar p p \rightarrow J/\psi + \gamma +X$ at the
Tevatron energy ($\sqrt{s} = 1.8$~TeV). The perturbative QCD
contributions to this process from the gluon-fusion
and the fragmentation mechanisms are computed. For the entire
range of $p_T$ that can be probed at the Tevatron,
the fusion contribution is found to be dominant.  Consequently the
QCD prediction for this process has better precision than the
purely hadronic production of $J/\psi$.  An experimental study
of this process at the Tevatron will help to clarify several aspects
of quarkonium production at large $p_T$.
\vspace{40pt}
\noindent
\begin{flushleft}
CERN-TH.7371/94\\
July 1994\\
\vspace{11pt}
$^{*)}$ dproy@theory.tifr.res.in \\
$^{**)}$ sridhar@vxcern.cern.ch\\
\end{flushleft}

\vfill
\clearpage
\setcounter{page}{1}
\pagestyle{plain}
The production of quarkonia at large $p_T$ is usually described by the
parton-fusion mechanism
in the framework of the colour-singlet model \cite{berjon,br,emc}.
In this model, the wave-function at the origin for a $S$-wave (or
its derivative for a $P$-wave) quarkonium is convoluted
with the cross-section for producing a heavy quark pair with
the proper spin, parity and charge-conjugation assignments.  The
wave-function $|R_0(0)|^2$ can be estimated from the $S$-wave
quarkonium decay widths within an uncertainty of $\sim 30$\%, arising
from relativistic correction \cite{nmc,schuler}.  The corresponding
uncertainty in the estimate of the derivative of the wave-function
$|R'_1 (0)|^2$ from $P$-wave quarkonium decay is much larger, since
the NLO correction for this case is not yet available \cite{nmc,schuler,
br2}.  This uncertainty propagates into the hadroproduction
cross-section of $J/\psi$, which is dominated by the $\chi \rightarrow
\psi + \gamma$ decay contributions.  Together with the uncertainty
from the choice of QCD scale in the $c\bar c$ pair production, one can
predict the $J/\psi$ hadroproduction cross-section to within a factor
of $\sim 3$.  Within this normalisation uncertainty, the colour
singlet model prediction agrees with the $J/\psi$ hadroproduction data
from fixed-target and ISR experiments over a reasonable range of $p_T$
\cite{br,schuler,isr,gms}.  A comprehensive account of the model
prediction and comparison with these data is given in \cite{schuler}.

\vskip10pt
At higher energies where $b$-quark production becomes important,
the decay of $b$ quarks is a mechanism that contributes to $J/\psi$
production, in addition to the parton-fusion mechanism of the
colour-singlet model \cite{nmc}. Recently, the CDF experiment studied
the large-$p_T$ $J/\psi$ production at the Tevatron collider
\cite{cdf} and, using the secondary
vertex information, separated out the $b$-quark decay
contribution. The remaining $J/\psi$'s are presumably produced directly
by the fusion mechanism, but the CDF experiment found that the
cross-section is an order of magnitude larger than that predicted
by the fusion mechanism of the colour-singlet model. Even if one
takes into account the above mentioned normalisation uncertainty in the
theoretical predictions, it does not suffice to restore agreement with
the CDF data.

\vskip10pt
The anomalously large cross-section measured in the
CDF experiment~-- the CDF effect~-- indicates that there is a new
mechanism through which $J/\psi$ production takes place at large $p_T$.
At high-energy colliders, where gluons and charm quarks are copiously
produced, the fragmentation of these particles into $J/\psi$
has been suggested as an important mechanism for large-$p_T$ $J/\psi$
production \cite{bryu}. The validity of this suggestion is borne
out by the explicit computation of the fusion and fragmentation
contributions for the Tevatron energy, presented in Refs.~\cite{ours,
bdfm,cg}. In these papers, it was shown that the fragmentation
contribution is the dominant contribution at the Tevatron energy; and
together with the fusion contribution it can explain the
large cross-section measured by the CDF experiment to within a factor
of 2-3. The largest contribution
comes from gluons fragmenting into $\chi$'s, which subsequently decay
into $J/\psi$'s. In
Ref.~\cite{ours}, the energy dependence of the fragmentation
and fusion contributions was studied as well. It was shown that at ISR
energies the fusion contribution dominates over the $p_T$ range of
interest, although the inclusion of the fragmentation contribution
helps to improve the quantitative agreement with the ISR data.

\vskip10pt
In Refs.~\cite{ours,bdfm}, the production of $\psi^{\prime}$ by
fragmentation was also studied. Unlike $J/\psi$ production,
$\psi^{\prime}$ production does not involve the $P$-state
contribution. Consequently, the fragmentation contribution is
comparatively smaller. The total predicted
$\psi^{\prime}$ cross-section is atleast an order of magnitude smaller
than the cross-section measured by CDF. This large discrepancy with
$\psi^{\prime}$ data seems to indicate a large contribution to the
$S$-wave quarkonium production from a still unknown mechanism. As
noted in \cite{ours,bdfm}, this will show up in a larger enhancement
of the $\psi^{\prime}$ cross-section than the $J/\psi$.  Whatever be
the source of this new contribution, it is important to probe it
phenomenologically through other processes of quarkonium production at
large $p_T$.  In this note, we point out that the production of a
large-$p_T$ $J/\psi$ with an associated photon will serve as a very
interesting probe for such mechanisms of quarkonium production.

\vskip10pt
As we shall see below, the associated production of a large-$p_T$
$J/\psi$ with a photon has three distinctive features in comparison
with the purely hadronic $J/\psi$ production discussed above.  1)~The
fusion contribution to this process dominates over the fragmentation
for the entire $p_T$ range of interest at the Tevatron energy.  2)~The
fusion contribution comes entirely from the direct production of the
$S$-state.  3)~The $S$-state wave function $|R_0 (0)|^2$ is known to
reasonable precision; and besides the uncertainty associated with the
QCD scale is much smaller here than the purely hadronic process.  In
short, the dominant contribution to this process comes from the direct
production of the $S$-state, which can be predicted to a reasonable
precision.  Thus it is well suited to probe for an anomalous
enhancement of the $S$-wave quarkonium production at large-$p_T$.

\vskip10pt
In recent years, there have been several discussions of associated
production of a $J/\psi$ and a photon in hadron-hadron experiments
\cite{drekim,psig1,psig2,psig3}.  This
production of a $J/\psi$--$\gamma$ pair can also take place at
HERA \cite{drekim,kimreya}, through resolved photon processes.
In the framework of the colour-singlet model, the production of a
$J/\psi$ and an associated photon, proceeds through the following
subprocess:
\begin{equation}
\label{e1}
g + g \rightarrow J/\psi + \gamma.
\end{equation}
$C$-invariance forbids a $\chi$ coupling to two gluons and a photon;
so a $\chi$ cannot be produced $via$ the above subprocess. Of course,
the $\chi$ contribution through $gg \rightarrow \chi \rightarrow
J/\psi + \gamma$ is possible; but this will contribute to
low-$p_T$ $J/\psi$'s and $\gamma$'s, and can be eliminated by a
$p_T$ or invariant mass cut. Therefore, large-$p_T$ $J/\psi$+$\gamma$
production through gluon fusion involves only the $S$-wave resonance.

\vskip10pt
In addition to the gluon-fusion contribution, one expects a
fragmentation contribution to the associated production of $J/\psi +
\gamma$ at large-$p_T$.  The fragmentation contribution to $J/\psi$ is
expected to come mainly from the gluon and charm-quark jets
\cite{bryu,ours,bdfm,cg}.  Thus the basic process of interest is the
associated
production of a large-$p_T$ photon with a gluon (or a charm quark),
followed by the fragmention of the latter into $J/\psi$ either
directly or via the $\chi$'s.  Consequently the fragmentation
contribution involves both $S$ and $P$ states.  Another distinctive
feature of the fragmentation contribution is the emergence of the
$J/\psi$ along with the other jet fragments.  Thus in principle one
could distinguish it from the fusion process (1) by requiring either
$J/\psi$ isolation or $p_T$ balancing between $J/\psi$ and $\gamma$.
It would be hard to implement this in practice, however, since the
$J/\psi$ is expected to carry away the bulk ($\sim 70$\%) of the jet
momentum \cite{bdfm}.  Therefore one has to deal with the sum of the
fusion and fragmentation contributions.

\vskip10pt
There is another source of $J/\psi + \gamma$ production at large-$p_T$
-- i.e. the associated production of a large-$p_T$ $J/\psi$ with a
hadron jet, followed by the fragmentation of the latter into a hard
photon \cite{koller,owens,aurenche}.  However this process can be
substantially reduced by the standard isolation cut on $\gamma$ used
in the direct photon experiments.  It was noted in \cite{aurenche}
that even with the isolation cut one should not neglect this
fragmentation contribution while dealing with the NLO correction to
direct photon production.  In the present work we shall work with the
lowest-order direct photon production cross-section and hence neglect
the contribution from photons produced via fragmentation.

\vskip10pt
Finally, we wish to make an observation here
concerning the model-dependence of the fusion contribution. We will
use the colour-singlet model of $J/\psi$ production in this work;
but there exists another model of $J/\psi$ production~-- the
semi-local duality model \cite{semi}. In this model, the $J/\psi$
cross-section is given by integrating the open $c \bar c$ cross-section
between $2m_c$ and the open $c \bar c$ production threshold and
multiplying by a normalisation factor, which is not specified by the
model but has to be determined by comparing with data. In this
model, the $c \bar c$ pair produced in the hard-scattering is not
required to be a colour-singlet, neither is a projection to the
quarkonium quantum numbers required. The colour and quantum numbers
are assumed to rearrange themselves appropriately, in the process
of soft-gluon emission. Consequently, in this model, the
$q \bar q$-fusion production of $J/\psi + \gamma$ is
allowed, in addition to the gluon-gluon fusion \cite{psig2}. However,
at the large energies of our present interest
the $gg$ fusion is expected to dominate in the duality model, so that the
predictions of the two models would not be significantly different.

\vskip10pt
The subprocess cross-section for $J/\psi$+$\gamma$ production in the
colour-singlet model is given as
\begin{eqnarray}
\label{e2}
&&{d\sigma \over d\hat t}(gg \rightarrow J/\psi + \gamma)
=  {16\pi\alpha\alpha_s^2 M \vert R_0(0) \vert^2
\over 27 \hat s^2} \biggl \lbrack
{\hat s^2 \over (\hat t -M^2)^2 (\hat u -M^2)^2}
\nonumber \\
&& + {\hat t^2 \over (\hat s -M^2)^2 (\hat u -M^2)^2}
+ {\hat u^2 \over (\hat t -M^2)^2 (\hat s -M^2)^2} \biggr \rbrack,
\end{eqnarray}
where $\hat s$, $\hat t$ and $\hat u$ are the usual Mandelstam
variables and $M$ is the mass of the $J/\psi$.
To obtain the hadronic $p_T$ distribution, this subprocess
cross-section is folded in with the gluon densities, as follows:
\begin{eqnarray}
\label{e3}
&&{d\sigma \over dp_T}(AB \rightarrow J/\psi+\gamma)
=  \nonumber \\
&& \sum \int dy \int dx_1 x_1g_{A}(x_1) x_2g_{B}(x_2)
{4p_T \over 2x_1 -\bar x_T e^y}
{d\hat \sigma \over d \hat t}(gg \rightarrow
J/\psi+\gamma) .
\end{eqnarray}
In the above equation, $g_A$ and $g_B$ are the gluon distributions
in $A$ and $B$, $x_1$ and $x_2$ are the respective momentum fractions.
Energy-momentum conservation determines $x_2$ to be
\begin{equation}
\label{e4}
x_2= {x_1 \bar x_T e^{-y} - 2 \tau \over 2x_1-\bar x_T e^y},
\end{equation}
where $\tau = M^2/s$, $\sqrt{s}$ the centre-of-mass energy, and $y$ the
rapidity at which the $J/\psi$ is produced. We also have
\begin{equation}
\label{e5}
\bar x_T= \sqrt{x_T^2 + 4\tau} \equiv {2M_T \over \sqrt{s}},
\hskip20pt x_T={2p_T \over \sqrt{s}} .
\end{equation}

\vskip10pt
As discussed above, in addition to the fusion process there
is a contribution to $J/\psi$+$\gamma$ production from the
fragmentation of gluons and charm quarks. At lowest order,
the subprocesses that give rise to a parton at large $p_T$ with
an associated photon are the usual direct photon subprocesses:
\begin{eqnarray}
\label{e6}
q + \bar q &\rightarrow g + \gamma \nonumber \\
Q + g &\rightarrow Q + \gamma ,
\end{eqnarray}
where we have used $Q$ to denote the charm quark, and the charm-quark
density in the initial state is that
generated by the $Q^2$ evolution of the structure
functions. Assuming factorisation,
the fragmentation contribution to the $J/\psi+\gamma$ cross-section
can be written as
\begin{equation}
\label{e7}
d\sigma (AB \rightarrow (J/\psi,\chi_i)+\gamma + X)
 = \sum \int_0^1 dz \hskip4pt
d\sigma (AB \rightarrow c +\gamma + X)
D_{c \rightarrow (J/\psi,\chi_i)}(z,\mu ) ,
\end{equation}
where $d\sigma (AB \rightarrow c +\gamma + X)$ is the hard-scattering
cross-section for the production of a parton (denoted by $c$) in
association with a photon, $D(z,\mu)$ is the fragmentation function
specifying the fragmentation of parton (gluon or charm quark)
into the required charmonium state, and $z$, as usual, is
the fraction of the momentum of the parent parton carried by the
charmonium state. The sum in the above equation runs over all
contributing partons. The fragmentation function is
computed perturbatively at an initial scale $\mu_0$,
which is of the order of $m_c$. It is then evolved to the scale
typical of the fragmenting parton, which is of the order of
$p_T/z$, using the Altarelli-Parisi equation.
The full set of initial fragmentation functions needed to
obtain the $J/\psi$ and the $\chi$ contributions have now been
computed. They are $D_{g \rightarrow J/\psi}$ \cite{bryu},
$D_{g \rightarrow \chi}$ \cite{bryu2}, $D_{c \rightarrow \psi}$
\cite{brcyu} and $D_{c \rightarrow \chi}$ \cite{chen,yuan}.

\vskip10pt
For the fragmentation process, the cross-section is given by a
formula similar to Eq.~\ref{e3} but with an extra integration
over $z$, or equivalently over $x_2$. We have
\begin{eqnarray}
\label{e8}
&&{d\sigma \over dp_T}(AB \rightarrow (J/\psi,\chi_i) +\gamma X)
=  \nonumber \\
&& \sum \int dy dx_1 dx_2 G_{a/A}(x_1) G_{b/B}(x_2)
D_{c\rightarrow (J/\psi,\chi_i)} (z)
{2p_T \over z} {d\hat \sigma \over d \hat t}(ab \rightarrow c+\gamma) ,
\end{eqnarray}
with $z$ given by
\begin{equation}
z= {\bar x_T \over 2} ({e^{-y} \over x_2} + {e^y \over x_1}) .
\end{equation}
For $d\hat\sigma/d\hat t(ab \rightarrow c +\gamma )$, we have used the
lowest-order direct photon cross-sections \cite{owens}.

\vskip10pt
In Fig.~1, we have presented our results for the
fusion and fragmentation contributions to $J/\psi + \gamma$ production
in $p \bar p$ collisions at $\sqrt{s}=1.8$~TeV.
We have plotted $Bd\sigma/dp_T$ as a function of $p_T$,
integrated over a pseudo-rapidity range $\vert \eta \vert < 0.5$,
where $B$ is the $J/\psi$ branching ratio into leptons ($B=0.0594$).
In our computations, we have used \cite{plothow} the updated
MRSD-${}^{\prime}$ parametrisations \cite{mrs} for the parton
densities in the nucleon.
The parton densities are evolved to a scale $Q^2=\mu^2/4$, where
$\mu$ is chosen to be $M_T$ for the fusion case,
and equal to $p_T^{g,c}=p_T/z$ for the fragmentation case.
The fragmentation functions are evolved to the scale $p_T/z$.
Another uncertainty that enters the normalisation of the predicted
cross-sections is that due to the wave-function at the
origin, $R_0$, the derivative for the $P$-states, $R_1^{\prime}$
(or equivalently $H_1$) and $H_8^{\prime}$, which is a parameter
that describes the $g \rightarrow \chi$ fragmentation through a
colour-octet mechanism.  The last parameter is related to the infrared
divergent term in the NLO correction to the $P$-state decay and has by
far the largest uncertainty. For the parameters $R_0$, $H_1$ and
$H_8^{\prime}$, we have used the values quoted in
Refs.~\cite{bryu,bryu2}: $R_0^2=0.8$~GeV${}^3$, $H_1=15.0$~MeV,
$H_8^{\prime}=3.0$~MeV.

\vskip10pt
The fusion contribution (shown as the solid line in Fig.~1) comes
entirely from the $S$-state production and hence depends only on
$R^2_0$. Being a gluon-gluon fusion process,
it is seen to dominate over the whole range of $p_T$ considered.
The gluon fragmentation contribution (including $g \rightarrow
J/\psi$ and $g \rightarrow \chi$ and shown as the dashed
line in Fig.~1) and the charm-quark fragmentation (shown as the
dotted line) are more than an order of magnitude smaller than the
fusion contribution, even at the largest values of $p_T$ considered.
These features may be contrasted with the purely hadronic $J/\psi$
production where the gluon fragmentation dominates over fusion and
both the contributions are dominated by the production of $P$-states
\cite{ours,bdfm}.

\vskip10pt
In Fig.~2, we show the $J/\psi$+$\gamma$ cross-section (i.e. the sum
of the fusion and fragmentation contributions) as a function of $p_T$.
This is shown as the solid line in Fig.~2.
The cross-section is large enough to be measurable at the Tevatron.
We also estimated the magnitude of the $\psi^{\prime}$+$\gamma$
cross-section, but it turns out to be two to three orders of
magnitude smaller than the $J/\psi$+$\gamma$ cross-section,
and may be too small to be observed. We have also considered the
effect of varying the scale from $\mu/2$
to $2\mu$ on the $J/\psi + \gamma$ cross-section (the latter choice is
shown as the dashed line in
Fig.~2). The cross-section is remarkably stable under scale variation.
This is due to a substantial compensation between the decrease of
$\alpha_s$ and the increase of the low-$x$ gluon distribution $g (x
\simeq .01)$ in eqs. 2,3 as the scale is increased from $\mu/2$ to
$2\mu$.  As a result the QCD prediction for the dominant process of
gluon fusion (1) is remarkably stable under scale variation at the
Tevatron energy.  The net uncertainty from the choice of scale as well
as the QCD parametrisation may be roughly estimated as $\pm 40$\%.  As
remarked before, the uncertainty from the estimate of the
corresponding wave function $R^2_0$ is also quite small ($\sim \pm
30$\%).  Combining the two gives an overall uncertainty of $\pm 50$\%
for the predicted $J/\psi + \gamma$ cross-section of Fig.~2.
Measurement of the cross-section at the Tevatron will therefore be
very useful in probing for any new source of quarkonium production at
large $p_T$.  In particular an anomalous enhancement of the $S$-wave
quarkonium production would show up here as clearly as in the purely
hadronic production of $\psi^{\prime}$ if it is associated with the
fusion mechanism, but not if it comes from the fragmentation.

\vskip10pt
In summary, we have analysed the associated production of a
large-$p_T$ $J/\psi$ and $\gamma$ at the Tevatron energy.  The
perturbative QCD predictions for the fusion and the fragmentation
contributions have been computed.  The fusion contribution is seen to
dominate this cross-section over the entire $p_T$ range of interest,
unlike the purely hadronic production of $J/\psi$.  More over the
fusion contribution to this process comes entirely from the direct
production of the $S$-wave quarkonium.  Consequently one has a rather
precise QCD prediction for this process.  Experimental measurement of
this cross-section will help to clarify several aspects of quarkonium
production at large $p_T$, and in particular those related to the
anomalous hadronic $J/\psi$ and $\psi^{\prime}$ cross-sections
recently reported by the CDF experiment (the CDF effect).

\vskip15pt
We would like to thank Drs. P. Sphicas and R.S. Bhalerao for
discussions.

\newpage
\section*{Figure captions}
\renewcommand{\labelenumi}{Fig. \arabic{enumi}}
\begin{enumerate}
\item
Lower figure: The cross-section $Bd\sigma/dp_T$ (integrated over the
pseudorapidity range $-0.5< \eta < 0.5$) for the process
$\bar p p \rightarrow J/\psi +\gamma + X$ as a function of $p_T$ at
$\sqrt{s}=1.8$~TeV. The different curves
correspond to the direct production via fusion (solid line), the
gluon fragmentation contribution (dashed line) and the charm-quark
fragmentation term (dotted line).

\item
The scale dependence of the $J/\psi + \gamma$ cross-section as a
function of $p_T$ for $\bar p p$ collisions at $\sqrt{s}=1.8$~TeV.
The solid curve in both figures is for the scale $\mu/2$
and the dashed curve is for the scale $2\mu$.

\end{enumerate}
\end{document}